\documentclass{article}
\usepackage{spconf,amsmath,graphicx}
\usepackage{multirow}
\usepackage{threeparttable}
\usepackage{setspace}
\usepackage{amssymb}
\usepackage{siunitx}
\usepackage{booktabs}
\usepackage{enumitem}

\title{Two-stream Joint-Training for Speaker independent acoustic-to-articulatory inversion}

\name{Jianrong Wang$^{1}$ 
\qquad Jinyu Liu$^{2}$ 
\qquad Li Liu$^{3\star}$
\qquad Xuewei Li$^{1}$ 
\qquad Mei Yu$^{1}$ 
\qquad Jie Gao$^{1}$
\qquad  Qiang Fang$^{4}$
\thanks{$^{\star}$Corresponding author: avrillliu@hkust-gz.edu.cn.}}
\address{$^{1}$ College of Intelligence and Computing, Tianjin University, Tianjin, China\\
$^{2}$ Tianjin International Engineering Institute, Tianjin University, Tianjin, China\\
$^{3}$ The Hong Kong University of Science and Technology (Guangzhou), Guangzhou, China\\
$^{4}$ Institute of Linguistics, Chinese Academy of Social Sciences, Beijing, China}

\begin{document}
%
\maketitle
\begin{abstract}

Acoustic-to-articulatory inversion (AAI) aims to estimate the parameters of articulators
from speech audio. There are two common challenges in AAI, which are the limited data and the unsatisfactory performance in speaker independent scenario. Most current works focus on extracting features directly from speech and ignoring the importance of phoneme information which may limit the performance of AAI. To this end, we propose a novel network called SPN that uses two different streams to carry out the AAI task. Firstly, to improve the performance of speaker-independent experiment, we propose a new phoneme stream network to estimate the articulatory parameters as the phoneme features. To the best of our knowledge, this is the first work that extracts the speaker-independent features from phonemes to improve the performance of AAI. Secondly, in order to better represent the speech information, we train a speech stream network to combine the local features and the global features. Compared with state-of-the-art (SOTA), the proposed method reduces 0.18mm on RMSE and increases 6.0\% on Pearson correlation coefficient in the speaker-independent experiment. The code has been released at https://github.com/liujinyu123/AAI-Network-SPN.
\end{abstract}
\vspace{-0.5cm}
\section{Introduction}
\label{sec:intro}
\vspace{-0.35cm}
Acoustic-to-articulatory inversion (AAI) \cite{C1} mainly solves the problem of deriving the pronunciation parameters of key organs from speech audio. In recent years, it has played an important role in many fields, such as pronunciation guidance \cite{C2} and speech recognition \cite{C4,C48,C50}, so it has attracted many researchers to devote themselves to this field.

Different deep learning based models and acoustic representations have been proposed to carry out the AAI task. In the early stage, codebook \cite{C7} was used for voice inversion, but the performance highly relied on the quality of codebook. Later, The data-driven voice inversion models were presented, such as hidden Markov model (HMM) \cite{C8,C49}, mixed Gaussian network (GMM) \cite{C38}, deep neural networks (DNNs) \cite{C39} and so on. At present, the most commonly used models are the recurrent neural network (RNN) and its variants, such as long-short term memory (LSTM) \cite{C14, C34}. In \cite{C14, C40, C41}, the different speech representations such as line spectral frequencies (LSF), Mel-frequency cepstral coefficients (MFCC) and filter bank energies (FBE) were used. In our work, we take MFCC and phonemes as the input of our model.

Up to now, there are two main challenges in AAI. One is that the available datasets are very limited, because we need to record the voice and pronunciation parameters at the same time, which is not only difficult to collect, but also expensive. The most commonly used public datasets are MOCHA-TIMIT \cite{C35}, MNGU0 \cite{C21}, and HPRC \cite{C36}. Another challenge is to improve the performance in speaker-independent scenarios.
\begin{figure*}[h]
	\centerline{\includegraphics[width=0.90\linewidth]{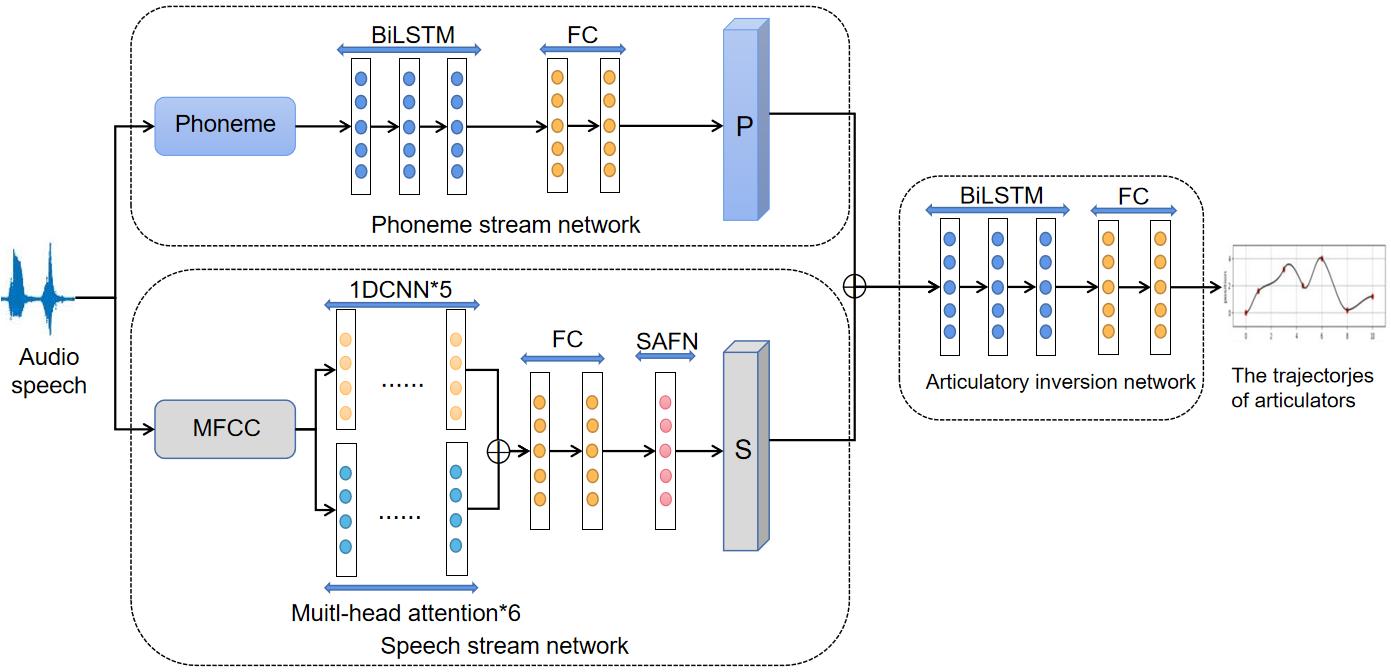}}
	\vspace{-0.3cm}
	\caption{The framework of SPN. The P is the phoneme stream feature and the S is the speech stream feature.}
	\label{fig:ov}
    \vspace{-0.4cm}
\end{figure*}

For the first challenge, \cite{C44} proposed the method of using the cross corpus data to solve the problem of limited data volume.  For the second challenge, \cite{C14} used the vocal tract normalization to map each speaker's pronunciation space to the same space, so as to improve the performance of that model, but it led to the loss of personalized information of speech. It is worth noting that a self-supervised pretraining model was proposed to solve the above challenges and achieved the best performance in \cite{C42}. However, this work only used MFCC as the network input which may limit the performance of AAI and used 1DCNNs to extract speech features which may result in the loss of global information of speech.   

In order to solve the above two challenges, we propose a novel network that consists of two parts, speech stream network and phoneme stream network, which we call SPN in brief. 1DCNNs were used to extract speech features in \cite{C42}. But it was pointed out that CNN only extracts the local features in \cite{C45}, so we add an multi-head attention model to extract the global features to better represent the voice information. In addition, we propose a new phoneme stream network. More precisely, we use transcribed phonemes to perform phoneme inversion, then take the results of phoneme stream network as the phoneme features to perform voice inversion. 
The motivation is that phonemes only represent the content information of the speech instead of the identity information. Therefore the phoneme features obtained by phoneme stream network are speaker-independent, which can improve the performance of the speaker-independent experiments.

In summary, there are three contributions of this work.
\begin{itemize}
\item In order to better represent voice information, we extract the local features and the global features through 1DCNNs and multi-head attention module respectively.
\item We propose a new phoneme stream network to gain the phoneme features to improve the performance of speaker-independent experiment.
\item Based on the experimental results, it is shown that the proposed model outperforms SOTA obviously on public HPRC dataset which decreases by 0.18mm on RMSE and increases by 6\% on PCC.
\end{itemize}
\vspace{-0.5cm}
\section{PROPOSED APPROACH}
\vspace{-0.25cm}
\label{sec:method}

\subsection{Overall Framework}
\vspace{-0.1cm}
As shown in Fig.~\ref{fig:ov}, the SPN we proposed is composed of two core modules, speech stream network and phoneme stream network. We get global and local features through speech stream network, then we feed them into the SAFN \cite{C42} network to obtain the speech features, and gain the phoneme features through phoneme stream network. Later, the integrated speech features and the phoneme features are fed to the articulatory inversion network (AIN) \cite{C42} to obtain the parameters of the key organs.

\subsection{Speech Stream Network}
\vspace{-0.1cm}
It was shown in \cite{C45} that CNN can only extract local features of speech, so to better represent voice information, a speech stream network is proposed. MFCC is fed into cascaded 1DCNNs and the multi-head attention module respectively to obtain the local features and the global features, which input into SAFN as speech features.
\begin{figure}[htb]
\begin{minipage}[b]{1.0\linewidth}
  \centering
  \centerline{\includegraphics[width=8.5cm,height=4cm]{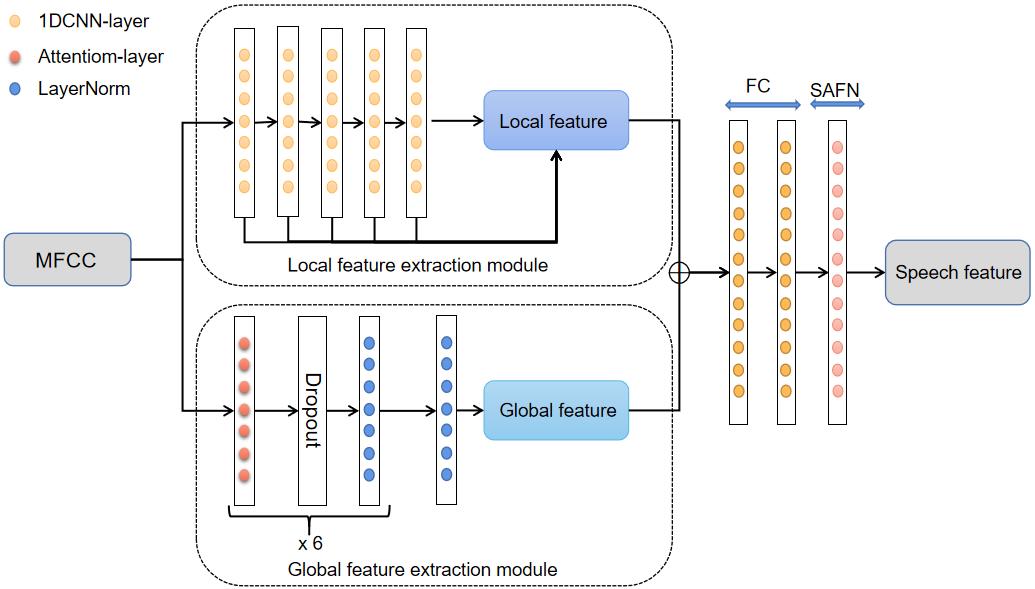}}
   \end{minipage}
  \caption{ The workflow of the Speech Stream Network.}
 \label{fig:FFN2}	
 \vspace{-0.4cm}
\end{figure}
As shown in Fig.~\ref{fig:FFN2}, the speech stream network is mainly divided into two parts, one is the local feature extraction module, the other is the global feature extraction module. In the local feature extraction module, we choose the cascaded 1DCNNs used in \cite{C42}. We send the MFCC into the 1DCNNs with five different convolution kernels, whose sizes are 1, 3, 5, 7 and 9, respectively, to obtain the local features. For the global feature extraction module, we use the multi-head attention module because it can pay attention to the relationship between each frame and other frames of the speech. The attention module can be described as obtaining the outputs from a set of queries, keys and values, where outputs, queries, keys and values are the corresponding vectors. Specifically, the global feature extraction module includes six layers of multi-head attention. Each layer has eight heads in total and the dimension of keys and values is both 64. Then, we feed the features into a layerNorm layer to get the global features. Next, the local features and global features are fed into two fully connected layers which has 300 units in each layer to get the speech features. Finally we feed the speech features to the SAFN. 

The computation of local features is formulated as:
\begin{equation}
\boldsymbol{y}_{i, j}^{\mathrm{local}}=b_{j}+\sum_{k=1}^{L_{i-1}} \mathbf{W}_{i} * \boldsymbol{y}_{i-1, k}^{\mathrm{local}}
\end{equation}
where * means the convolution operation and ${y}_{i, j}$ represents the feature map of $j$-th channel in $i$-th convolution layer, $b_{j}$ is the bias of $j$-th channel and $\mathbf{W}_{i}$ means the weights of $i$-th convolution layer, $L_{i-1}$ means the length of $\boldsymbol{y}_{i-1}^{\mathrm{local}}$.



\vspace{-0.3cm}
\subsection{Phoneme Stream Network}
\vspace{-0.1cm}
Inspired by the \cite{C43}, we use the outputs of phoneme stream network as the phoneme features to assist voice inversion. The reason for this is the following: phoneme itself only encodes the content information of the speech, instead of the identity information which is speaker-independent. We use the Penn phonetics lab forced aligner \cite{C46} to extract the phoneme frames. Each phoneme frame is a 39 dimensional one-hot vector.

In this module, we use three-layer BLSTM, and finally feed it to the two fully connected layers to get the pronunciation parameters as the phoneme features. The three-layer BLSTM has the same setting, which has 150 activation units in each layer, and there are 300 activation units for the fully connected layer. The output are the 12 dimensional pronunciation parameters, then we feed the pronunciation parameters as phoneme features and the speech features obtained from the speech stream network module into the AIN network to perform voice inversion. The core of the phoneme stream network is expressed as :

\begin{equation}
\setlength{\abovedisplayskip}{3pt}
s_{i} =f\left(U x_{i}+W s_{i-1}+b\right), \notag
\end{equation}
\begin{equation}
\setlength{\abovedisplayskip}{3pt}
s_{i}^{\prime} =f\left(U^{\prime} x_{i}+W^{\prime} s_{i+1}+b^{\prime}\right), 
\end{equation}
\begin{equation}
o_{i} =g\left(V s_{i}+V^{\prime} s_{i}^{\prime}+c\right),\notag
\end{equation}
where $o_{i}$ is the phoneme features estimated at the frame $i$. ${x}_{i}$ is the input phoneme sequence at frame $i$, $s_{i}$ is the temporary state at frame $i$ and $U, W, U^{\prime}, W^{\prime}$ are the corresponding transformation matrices. $b, b^{\prime}$  are the biases.


\vspace{-0.35cm}
\section{EXPERIMENTS}
\vspace{-0.25cm}
\subsection{Experimental Setup}
\vspace{-0.1cm}
\textbf{Dataset.} In this work, the dataset we used is HPRC. There are six locations: T1, T2, T3, UL, LL and LI. To be consistent with the SOTA, we only use the locations of tongue organs in X and Z directions as our experimental predicted labels (\textit{e.g}., T1, T2, T3). The HPRC \cite{C36} dataset has eight native American English speakers (four males and four females), and each speaker data is made up of 720 utterances.

\textbf{Performance Metrics.} The performance of the experiment is measured by mean square error and Pearson correlation coefficient \cite{C11}. The first indicator measures the error between the predicted label and the real label and the second indicator represents the similarity of the two labels. 


\textbf{Implementation Details.} The network based on speech decomposition and auxiliary feature proposed in \cite{C42} is the SOTA in AAI. Besides, the settings of SAFN and AIN modules in the network architecture diagram are the same as those of SOTA. In this work, we trained the model for 20 epochs and used the Adam optimizer. The learning rate is 1e-4, and the batch size is 5.
\vspace{-0.3cm}
\subsection{Comparisons with the SOTA}

To check the effectiveness of our proposed SPN network, we set three different experimental scenarios according to Table.~\ref{tab:TAB1}. S1 represents that we only train the phoneme stream network using phonemes as input to conduct speaker-independent experiments. S2 represents that we take phonemes and MFCC as inputs to conduct speaker-independent experiments. In this scenario, we take phoneme stream network as a pretraining model, and freeze the parameters while training the whole SPN. S3 represents speaker-independent experiments with phonemes and MFCC as network inputs. Unlike S2, we train phoneme stream network with the whole network. The loss function of our network is the weighted sum of the two parts which are the L2 loss of SPN and L2 loss of phoneme stream network. The formula is given as: 


\begin{equation}
\setlength{\abovedisplayskip}{3pt}
\setlength{\belowdisplayskip}{3pt}
	{L}_{joint} = \sum_{i=0}^{k}\left(y_{m}^{i}-\hat{y}_{m}^{i} \right)^{2} + \sum_{i=0}^{k}\left(y_{n}^{i}-\hat{y}_{n}^{i} \right)^{2},
\end{equation} 
where $\hat{y}_{m}^{i}$, $\hat{y}_{n}^{i}$ represent the ground-truth of EMA. ${y}_{m}^{i}$, ${y}_{n}^{i}$ are the corresponding predicted labels. $k$ means the length of the speech feature.

More specifically, we calculated the performance of phoneme stream network and SPN at the same time in Scene 3. S3(P) represents the result of phoneme stream network and S3(S) represents the results of SPN.
It is worth noting that our experiment was conducted on the Haskins dataset and in each scenario, we trained a separate model for each speaker. We collect all the training set (80\%), validation set (20\%) from seven speakers' data and test on the left one speaker data (100\%). Finally, we take the average of the results of the eight speakers as the final results of our experiments.

\begin{table}[h]
	\caption{ Experimental setup for 3 different scenarios. * means taking the proportion of data from each speaker. P and M represent phoneme and MFCC, respectively. }
	\label{tab:TAB1}
	\centering 
	\begin{threeparttable}
		\scriptsize
			\setlength{\tabcolsep}{1.8mm}{
				\begin{tabular}{ccccccc}
					\toprule
					\textbf{Scenarios}&\textbf{Input}&\textbf{\#Speaker}&\textbf{Train}&\textbf{Validation}&\textbf{Test}\\
					\midrule
					\multirow{2}{*}{\textbf{S1}}&\multirow{2}{*}{P}&N-1&80\%*&20\%*&-{}-{}-\\
					&&1&-{}-{}-&-{}-{}-&100\%\\
                    \midrule
                    \multirow{2}{*}{\textbf{S2}}&\multirow{2}{*}{P, M}&N-1&80\%*&20\%*&-{}-{}-\\
					&&1&-{}-{}-&-{}-{}-&100\%\\
                    \midrule
					\multirow{2}{*}{\textbf{S3}}&\multirow{2}{*}{P, M}&N-1&80\%*&20\%*&-{}-{}-\\
					&&1&-{}-{}-&-{}-{}-&100\%\\
					\bottomrule
			\end{tabular}}
	\end{threeparttable}
\end{table}

\begin{table}[h]
	\centering
    \vspace{-0.4cm}
	\begin{threeparttable}
		\caption{RMSE and PCC for SPN and SOTA}
		\label{tab:TAB3}
		\centering
		\scriptsize
		\begin{spacing}{0.85}
			\setlength{\tabcolsep}{0.8mm}{
				\begin{tabular}{ccccccccccc}
					\toprule
					\textbf{Scenarios}&\textbf{F01}&\textbf{F02}&\textbf{F03}&\textbf{F04}&\textbf{M01}&\textbf{M02}&\textbf{M03}&\textbf{M04}&\textbf{RMSE}&\textbf{PCC} \\
					\midrule
					{\textbf{S1}}&2.279&2.182&1.681&2.322&1.753&2.546&2.036&1.898&2.087&0.755\\
					\midrule
					{\textbf{S2}}&2.720&2.740&2.088&2.759&2.352&3.064&2.395&2.528&2.580&0.798\\
					
					\midrule
					{\textbf{S3(P)}}&2.169&2.084&1.618&2.172&1.693&2.440&1.961&1.812&1.993&0.773\\
					
					\midrule
					{\textbf{S3(S)}}&2.634&2.665&2.100&2.602&2.235&3.195&2.368&2.504&2.537&0.810\\
					\midrule
					{\textbf{SOTA}}&-&-&-&-&-&-&-&-&2.721&0.751\\
                    \midrule
                    
			\end{tabular}}
		\end{spacing}
	\end{threeparttable}
    \vspace{-0.5cm}
\end{table}

\begin{figure}[h]
	\centering
	\includegraphics[width=0.8\linewidth]{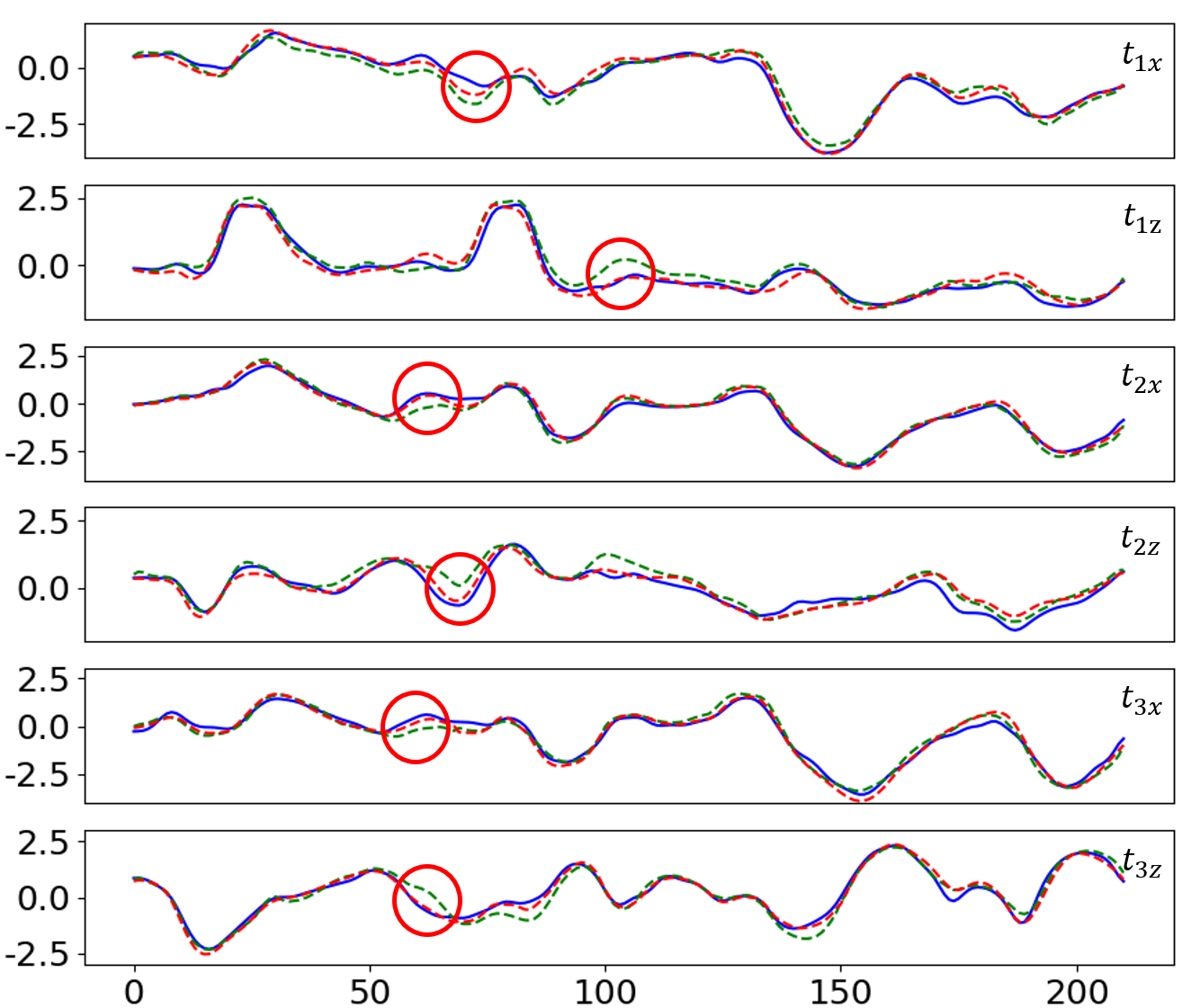}
	\caption{Comparisons of the predicted results of the tongue movement between SPN and SOTA. In each box, the blue line is real label, the red dashed line is predicted by SPN and the green dashed line is predicted by SOTA method.}
	\label{fig:draw1} 
    \vspace{-0.4cm}
\end{figure}

Table.~\ref{tab:TAB3} shows RMSE and PCC value in three scenarios in Haskins dataset. Because the previous SOTA works did not give the average RMSE on each speaker, we set the corresponding positions as '-' in Table.~\ref{tab:TAB3}. Basically, we can clearly observe that compared with the SOTA, our proposed SPN network decreases 0.141mm on RMSE and increases almost 5\% on PCC in scenario 2, decreases 0.184mm on RMSE and increases almost 6\% on PCC in scenario 3 which uses the joint-training strategy. It shows that the local features and global features we extracted can better represent speech information. After adding the phoneme features, we can further effectively improve the generalization ability of this model.

More interestingly, from the experimental results, we can see that in the case of scenario 3, the performance of phoneme stream network under joint training (S3(P)) is better than that training alone (S1), about 0.1mm on RMSE and almost 2\% on PCC. This shows that the performance of SPN and phoneme stream network both improved, indicating the effectiveness of joint-training strategy.

 Qualitative comparisons between SPN and SOTA are shown in Fig.~\ref{fig:draw1}. It is obvious that the predicted tongue articulators generated by SPN are more similar to ground-truth than the SOTA, especially in the red circle of the Fig.~\ref{fig:draw1}. It indicates that the phoneme stream network and speech stream network have a positive effect on the reconstruction of tongue articulatory movements.
\vspace{-0.25cm}
\subsection{Ablation Study}
\vspace{-0.1cm}
To prove the effectiveness of the proposed models, we conduct the ablation experiment according to Table.~\ref{tab:TAB2}, and the results are presented in Table.~\ref{tab:TAB4}.
\begin{table}[h]
	\scriptsize
    \vspace{-0.5cm}
	\caption{\textbf\tiny{Ablation experiment proves the effectiveness of each module. SPN-S represents the model that combines the speech stream network and SOTA model.}}
	\centering
	\label{tab:TAB2} 
	\begin{threeparttable}
		\begin{spacing}{0.8}
			\setlength{\tabcolsep}{5.0mm}{
				\begin{tabular}{cccc}
					\toprule
					\textbf{}&\textbf{speech stream network}&\textbf{phoneme stream network}\\
					\midrule
					\textbf{SOTA}&\textbf{$\times$}&\textbf{$\times$}\\
					\textbf{SPN-S}&\textbf{\checkmark}&\textbf{$\times$}\\
					\textbf{SPN}&\textbf{\checkmark}&\textbf{\checkmark}\\
					\bottomrule
			\end{tabular}}
		\end{spacing}
	\end{threeparttable}
\end{table}

Obviously, the models with speech stream network (SPN, SPN-S) outperform that without the speech stream network (SOTA). Besides, the model with phoneme stream network (SPN) outperforms that without phoneme stream network (SPN-S). From the experimental results, we can clearly observe that in the speaker-independent experiment, the speech stream network improves the performance by extracting the local features and global features to better represent the speech information and after adding the phoneme features obtained by phoneme stream network brings a further large gain on the generalization ability of our model.

\begin{table}[h]
	\scriptsize
    \vspace{-0.35cm}
	\caption{\textbf\tiny{RMSE and CC in three scenarios.}}
	\centering
	\label{tab:TAB4} 
	\begin{threeparttable}
		\begin{spacing}{0.8}
			\setlength{\tabcolsep}{5.0mm}{
				\begin{tabular}{cccc}
					\toprule
					\textbf{}&\textbf{RMSE}&\textbf{PCC}\\
					\midrule
					\textbf{SOTA}&\textbf{2.721}&\textbf{0.751}\\
					\textbf{SPN-S}&\textbf{2.664}&\textbf{0.787}\\
					\textbf{SPN}&\textbf{2.537}&\textbf{0.810}\\
					\bottomrule
			\end{tabular}}
		\end{spacing}
	\end{threeparttable}
\end{table}

\vspace{-0.7cm}
\section{Conclusion}
\vspace{-0.2cm}
To improve the speaker-independent performance and pay more attention to the global speech information, we propose a new network, including two parts. One is the speech stream network, which uses 1DCNNs and multi-head attention model to extract the local features and global features of speech to better represent the voice information. At the same time, we use the pronunciation parameters obtained by the phoneme stream network as the phoneme features to help predict the pronunciation parameters of voice inversion. The experimental results prove the effectiveness of our proposed network. In the future work, meta-learning will be explored into the field of AAI.
\vspace{-0.4cm}
\section{Acknowledgement}
\vspace{-0.2cm}
This work is supported by the National Natural Science
Foundation of China (No. 61977049), the National Natural Science Foundation of China (No. 62101351), and the GuangDong Basic and Applied Basic Research Foundation
(No.2020A1515110376).

\bibliographystyle{IEEEbib}
\bibliography{mybib}
\end{document}